# A Pseudo 2D-analytical Model of Dual Material Gate All-Around Nanowire Tunneling FET

Rajat Vishnoi and M. Jagadesh Kumar, *Senior Member, IEEE*

*Abstract—* In this paper, we have worked out a pseudo two dimensional (2D) analytical model for surface potential and drain current of a long channel p-type Dual Material Gate (DMG) Gate All-Around (GAA) nanowire Tunneling Field Effect Transistor (TFET). The model incorporates the effect of drain voltage, gate metal work functions, thickness of oxide and silicon nanowire radius. The model does not assume a fully depleted channel. With the help of this model we have demonstrated the accumulation of charge at the interface of the two gates. The accuracy of the model is tested using the 3D device simulator Silvaco Atlas.

## I. INTRODUCTION

Studies on novel device structures for VLSI applications is being done extensively these days to provide alternatives to conventional CMOS transistors. This is because MOSFETs scaled to lengths below 100 nm face several problems with regard to leakage currents in the OFF-state, subthreshold swing (SS), drain induced barrier lowering (DIBL) and numerous other short channel effects (SCE). An attractive alternative is TFET [1-6], which has SS below 60 mV/decade, low OFF-state leakage currents and diminished short channel effects (SCE). However, since the source of carriers is band to band tunneling, TFETs have a low ON-state current $I_{ON}$ [3] and it does not meet ITRS requirements [7-9]. Also, a gate all around (GAA) structure provides an improved SS and a solution to SCE and DIBL [10-12] because of the better electrostatic control over the channel. Further, a higher $I_{ON}$ per unit device area is achieved due to the device geometry [13-15]. However, the problems of delayed saturation, low $I_{ON}$ and leakage currents remain in a GAA structure. Thus, to optimize the GAA structure, a Dual Material Gate (DMG) structure is proposed in [16]. A DMG GAA nanowire TFET (Fig. 1) has a tunneling gate with a work function lower than that of the auxiliary gate for an n-channel TFET and vice-versa for a p-channel TFET. The DMG GAA nanowire TFET will provide a higher $I_{ON}$ due to the increased tunneling by a metal of lower work function. It will have a lower OFF-state current ($I_{OFF}$) because of the presence of a minimum in the surface potential and a negative

The authors are with the Department of Electrical Engineering, Indian Institute of Technology, Delhi, 110 016 India (e-mail: vishnoir@gmail.com; mamidala@ee.iitd.ac.in).

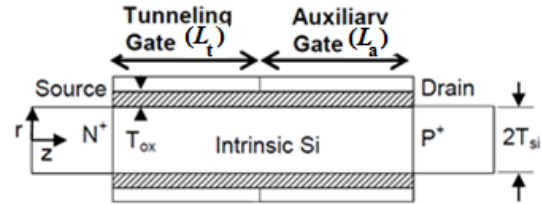

Fig 1. Schematic view of a cross-section of the p-channel DMG GAA nanowire TFET.

electric field in the channel. Thus we will get a better $I_{ON}/I_{OFF}$ ratio and a better SS [3, 15, 16]. Thus, a DMG GAA nanowire TFET provides a higher $I_{ON}$ with reduced SCEs and better switching characteristics as compared to a planar TFET. The DMG structure has been explored extensively in literature [17-20]. Although, TFETs using other materials are being studied, there is a great interest in silicon TFETs with improved $I_{ON}$ because of their suitability in silicon based CMOS technology.

The DMG GAA nanowire TFET has been studied in [15] using numerical simulations but a compact model for drain current characteristics is needed for a better understanding of the working of the device. Several models have been developed for conventional TFETs [21-26] but a model for DMG GAA TFET is yet to be developed. The objective of this work is, therefore, to develop a pseudo-2D analytical model for the DMG GAA nanowire TFET using the approaches in [27] and [28].

In this work, we begin with the modeling of the surface potential along the channel using a pseudo-2D model [27] for solving the Poisson equation in the silicon channel. We then find the average electric field in the tunneling region and derive the drain current using Kane's model [29]. To begin with, we develop a model considering equal lengths of tunneling and auxiliary gates but we later extend our model to a more general case where the tunneling gate length is much smaller than the auxiliary gate length as suggested in [16] for optimal results. The accuracy of our models is validated by comparing the results given by our models with three dimensional numerical simulations [30]. The tunneling parameters for the simulations are calibrated by accurately reproducing experimental results published in Fig 6 (a) of [28].

## II. MODEL DERIVATION

The schematic view of the p-channel DMG GAA nanowire TFET structure is shown in Fig. 1. The length of the channel is 200 nm and the length of the source and drain regions is 50 nm.

The channel doping $N_S = 10^{15}/cm^3$ and the source and drain dopings are $10^{21}/cm^3$. The radius of the nanowire ($T_{Si}$) is 10 nm and the oxide thickness ($T_{ox}$) is 2 nm. The gate which surrounds the entire channel is split into two parts. Tunneling gate at the source side with a work function $\Phi_{tunnel} = 4.8$ eV and an auxiliary gate at the drain side with $\Phi_{aux} = 4.4$ eV as suggested in [16] for optimal results. As a general case we begin with equal lengths of both the gates i.e. tunneling gate length $L_t = 100$ nm and auxiliary gate length $L_a = 100$ nm and later extend it for a tunneling gate of shorter length i.e. $L_t = 20$ nm and $L_a = 180$ nm as suggested by [16] for optimal results.

Fig. 2 shows the simulated band diagram and the surface potential of the device along the z-direction. As the current in a TFET is low, it can be observed that the potential drop along the channel is extremely small in the regions shown by the solid arrows in the figure and can be assumed to be constant [28]. Hence it can be inferred that the channel is not depleted in these regions. In Fig. 3, the surface potential of a DMG GAA nanowire TFET has been compared with that of two SMG GAA nanowire TFETs with gate work functions $\Phi_{SMG} = 4.4$ eV and 4.8 eV. As can be seen here, in a DMG GAA nanowire TFET the two non-depleted regions in the channel, one under each gate, has a potential equal to that in an SMG GAA nanowire TFET of the corresponding gate work function. The values of these constant regions under the tunneling and the auxiliary gates are denoted as $\psi_{Ct}$ and $\psi_{Ca}$, respectively.

In the channel, TFETs behave like regular MOSFETs apart from the tunneling region at the source end. This is because the mechanism of channel formation and charge transport is the same in both TFETs and MOSFETs. Therefore, the value of $\psi_{Ct}$ is given by

$$\psi_{Ct} = V_D + \psi_{Bt} \qquad \text{if } |V_D| \le |V_G - V_{tha}| \qquad (1)$$

$$\psi_{Ct} = V_{GS} - V_{tha} + \psi_{Bt} \qquad \text{if } |V_D| \ge |V_G - V_{tha}| \qquad (2)$$

where $\psi_{Bt}$ is the channel's built-in potential under the tunneling gate due to the band bending caused by the gate voltage, and $V_{tha}$ is the threshold voltage of a MOSFET with a gate work function $\Phi = \Phi_{aux}$. The expression for equation (2) will be different for a DMG GAA nanowire TFET with $\Phi_{aux} > \Phi_{tunnel}$. In this case, $V_{tha}$ would be replaced by $V_{tht}$ which is the threshold voltage for a MOSFET with a gate work function $\Phi = \Phi_{tunnel}$. This happens because for a pTFET when $\Phi_{aux} < \Phi_{tunnel}$ then $|V_{tha}| > |V_{tht}|$ and hence, with increasing $V_D$ saturation happens in the auxiliary channel first and the entire channel potential gets saturated. Whereas, when $\Phi_{aux} > \Phi_{tunnel}$, the channel under the tunneling gate gets saturated first and the auxiliary channel potential is dependent on the drain voltage until it gets saturated. The value of $\psi_{Ca}$ is given by

$$\psi_{Ca} = V_D + \psi_{Ba} \qquad \text{if } |V_D| \le |V_G - V_{tha}| \qquad (3)$$

$$\psi_{Ca} = V_{GS} - V_{tha} + \psi_{Ba} \qquad \text{if } |V_D| \ge |V_G - V_{tha}| \qquad (4)$$

where $\psi_{Ba}$ is the channel's built-in potential under the auxiliary gate due to the band bending caused by the gate voltage. Also,

$$\psi_{Source} = V_S + V_{bi} \qquad (5)$$

$$\psi_{Drain} = V_D \qquad (6)$$

where $V_{bi}$ is the built-in potential of the source-body junction and $V_S$ is at ground potential.

The channel has four depletion regions, R1, R2, R3 and R4 in saturation as shown in Fig. 2 and only R1, R2 and R3 in linear region. Depletion region R4 need not be considered for drain current calculations, since surface potential near source region is only needed. The surface potential in these regions can be modelled by solving the 2D Poisson equation in cylindrical coordinates as given by:

$$\frac{1}{r}\frac{\partial}{\partial r}\left(r\frac{\partial \psi(r,z)}{\partial r}\right) + \frac{\partial^2 \psi(r,z)}{\partial z^2} = \frac{qN_S}{\varepsilon_{Si}} \qquad (7)$$

Fig. 4 shows the potential profile along the $r$-direction of a cross-section of the DMG GAA nanowire TFET. The shape of this profile can be approximated by a second order polynomial [27]:

$$\psi(r,z) = a_0(z) + a_1(z)r + a_2(z)r^2 \qquad (8)$$

This polynomial has to satisfy the following three boundary conditions as can be seen in Fig. 4. The potential $\psi$ at $r = T_{Si}$ equals the surface potential $\psi_S$:

$$\psi(T_{Si}, z) = \psi_S(z) \qquad (9)$$

Electric field at $r = 0$ is equal to zero:

$$E_r(0, z) = 0 \qquad (10)$$

Electric displacement field at $r = T_{Si}$ is equal across the silicon and oxide boundary:

$$E_r(T_{Si}, z) = -C_{ox}(\psi_G - \psi_S(z))/\varepsilon_{Si} \qquad (11)$$

where $\psi_G$ is the electrostatic potential of the gate and is equal to $V_G - V_{FB}$ ($V_{FB}$ is the flat band voltage for the tunneling gate) and $C_{ox}$ is oxide capacitance per unit area at $r = T_{Si}$, given by

$$C_{ox} = \varepsilon_{ox}/(T_{Si}\ln(1+\frac{T_{ox}}{T_{Si}})) \qquad (12)$$

By applying these boundary conditions to (8), we get:

$$a_0(z) = \psi_S(z)(1+\frac{T_{Si}^2 C_{ox}}{2T_{Si}\varepsilon_{Si}}) - \frac{T_{Si}^2 C_{ox}}{2T_{Si}\varepsilon_{Si}}\psi_G \qquad (13)$$

$$a_1(z) = 0 \qquad (14)$$

$$a_2(z) = \frac{C_{ox}}{2T_{Si}\varepsilon_{Si}}(\psi_G - \psi_S(z)) \qquad (15)$$

Using (8) in (7), the surface potential can be written as:

$$\frac{\partial^2 \psi_S}{\partial z^2} - \frac{2C_{ox}}{T_{Si}\varepsilon_{Si}}\psi_S = \frac{qN_S}{\varepsilon_{Si}} - \frac{2C_{ox}}{T_{Si}\varepsilon_{Si}}\psi_G \qquad (16)$$

which gives the following general form solution for the surface potential in regions R1, R2 and R3.

$$\psi_{si}(z) = C_i \exp(\frac{z-L_i}{L_d}) + D_i \exp(\frac{-(z-L_i)}{L_d}) + \psi_{Gi} - \frac{qN_i L_d^2}{\varepsilon_{Si}} \qquad (17)$$

$$L_d = \sqrt{T_{Si}^2 \ln(1+\frac{T_{ox}}{T_{Si}})\varepsilon_{Si}/(2\varepsilon_{ox})} \qquad (18)$$

where $C_i$ and $D_i$ are unknown coefficients, $L_i$ is the length of region $R_i$, $N_i$ is the background negative charge concentration of region $R_i$, $\psi_{Gi}$ is the electrostatic potential gate over region $R_i$ and $L_d$ is the characteristic length. In region R1, (17)

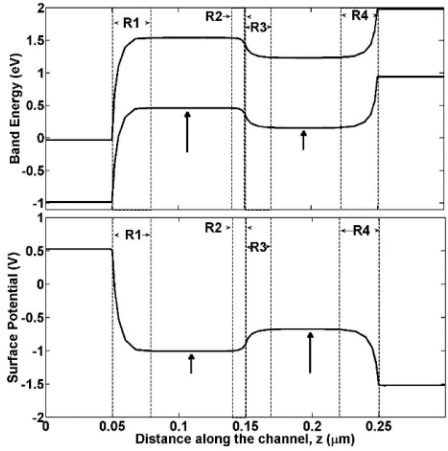

Fig 2. Simulated band diagram (upper curve) and surface potential (lower curve) for the DMG GAA nanowire TFET with $\Phi_{tunnel}$ = 4.8 eV and $\Phi_{aux}$ = 4.4 eV at $V_{GS}$= -1.0 V and $V_{DS}$ = -1.0 V. The depletion regions are marked by R1, R2, R3 and R4 and the non-depleted regions are shown by solid arrows.

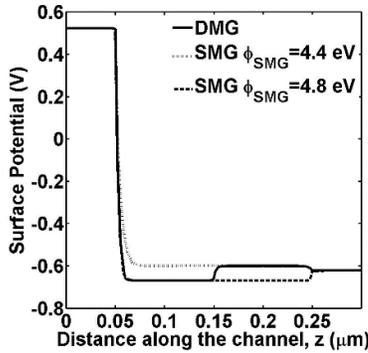

Fig 3. Simulated surface potential profiles of the DMG GAA nanowire TFET compared with that of two SMG GAA nanowire TFETs having gate work functions $\Phi_{SMG}$ = 4.4 eV and 4.8 eV at $V_{GS}$ = -1.0 V and $V_{DS}$ = -0.1 V.

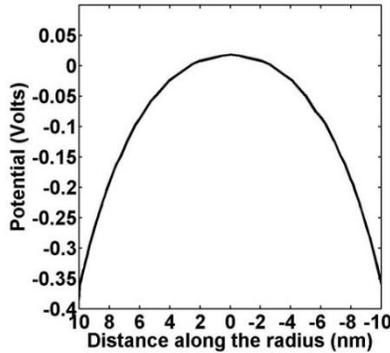

Fig 4. Potential distribution in the channel along the radial direction in region R1 for $V_{GS}$ = -1.0 V and $V_{DS}$ = -0.1 V.

has to satisfy the following boundary conditions (defining z = 0 at source-channel boundary):

(a) Surface potential at source is equal to $V_{bi}$:

$$\psi_{s1}(0) = V_{bi} \qquad (19)$$

(b) $\psi_{S1}$ at the boundary of region R1 is equal to $\psi_{Ct}$:

$$\psi_{s1}(L_1) = \psi_{Ct} \qquad (20)$$

(c) Electric field at the boundary of region R1 is equal to zero:

$$\frac{\partial \psi_{s1}(L_1)}{\partial z} = 0 \qquad (21)$$

From (20) and (21), we get the following expression for $\psi_{S1}$:

$$\psi_{s1}(z) = (\psi_{Ct} - \psi_{Gt} + \frac{qN_S L_d^2}{\varepsilon_{Si}})\cosh(\frac{z - L_1}{L_d}) + \psi_{Gt} - \frac{qN_i L_d^2}{\varepsilon_{Si}} \qquad (22)$$

where $L_1$ is the length of R1 and can be evaluated by using (19).

Let us now model the surface potential in regions R2 and R3. One important thing to note here is that the value of $N_i$ will be different for R2 and R3. The channel at the boundary between R2 and R3 behaves like a $p^+$-p junction and hence there will be complete depletion in R2 but the mobile charges of the channel which were earlier in region R2 will now move into region R3. We assume this charge in region R3 to be $n$ per cm³. Hence, $N_i$ in R2 will be the body doping $N_S$ and in R3, it will be $N_S + n$. There will be two equations like (17) one each for regions R2 and R3. We will have six unknown parameters ($C_2$, $D_2$, $L_2$, $C_3$, $D_3$, and $L_3$) which can be determined by the following boundary conditions assuming that z = 0 is the junction of the two gates.

(i) The value of $\psi_S$ at z = -$L_2$ is equal to $\psi_{Ct}$ and at z = $L_3$ is equal to $\psi_{Ca}$.

$$\psi_{s2}(-L_2) = \psi_{Ct} \qquad (23)$$

$$\psi_{s3}(L_3) = \psi_{Ca} \qquad (24)$$

(ii) The electric field at z = -$L_2$ and at z = $L_3$ is zero:

$$\frac{\partial \psi_{s2}(-L_2)}{\partial z} = 0 \qquad (25)$$

$$\frac{\partial \psi_{s3}(L_3)}{\partial z} = 0 \qquad (26)$$

(iii) The surface potential is continuous at z = 0:

$$\psi_{s2}(0) = \psi_{s3}(0) \qquad (27)$$

(iv) The electric field is continuous at z = 0:

$$\frac{\partial \psi_{s2}(0)}{\partial z} = \frac{\partial \psi_{s3}(0)}{\partial z} \qquad (28)$$

From (23) and (25), we get

$$C_2 = D_2 = \frac{\psi_{Ct} - \psi_{Gt} + \frac{qN_S L_d^2}{\varepsilon_{Si}}}{2} \qquad (29)$$

From (24) and (26), we get

$$C_3 = D_3 = \frac{\psi_{Ca} - \psi_{Ga} + \frac{q(N_S + n)L_d^2}{\varepsilon_{Si}}}{2} \qquad (30)$$

From (27), we get

$$\frac{\psi_{Ct} - \psi_{Gt} + \frac{qN_S L_d^2}{\varepsilon_{Si}}}{2} \times (e^{\frac{L_2}{L_d}} + e^{\frac{-L_2}{L_d}}) + \psi_{Gt} - \frac{qN_S L_d^2}{\varepsilon_{Si}} =$$
$$\frac{\psi_{Ca} - \psi_{Ga} + \frac{q(N_S + n)L_d^2}{\varepsilon_{Si}}}{2} \times (e^{\frac{L_3}{L_d}} + e^{\frac{-L_3}{L_d}}) + \psi_{Ga} - \frac{q(N_S + n)L_d^2}{\varepsilon_{Si}} \qquad (31)$$

From (28), we get

$$\frac{\psi_{Ct} - \psi_{Gt} + \frac{qN_S L_d^2}{\varepsilon_{Si}}}{2} \times (e^{\frac{L_2}{L_d}} - e^{\frac{-L_2}{L_d}}) = \frac{\psi_{Ca} - \psi_{Ga} + \frac{q(N_S + n)L_d^2}{\varepsilon_{Si}}}{2} \times (e^{\frac{-L_3}{L_d}} - e^{\frac{L_3}{L_d}})$$
$$\qquad (32)$$

The value of $n$ in region R3 can be calculated by adding the following condition in our model, which comes from the conservation of charge across the p⁺-p junction at the boundary of the tunneling gate and the auxiliary gate i.e.

$$nL_3 = n_{ch2}L_2 \qquad (33)$$

where $n_{ch2}$ is the channel inversion charge concentration under the tunneling gate. Simultaneously solving (31), (32) and (33) gives us the values of $L_2$, $L_3$ and n. These equations have to be solved numerically as they contain non-linear expressions. Since, after the onset of strong inversion, the channel charge does not vary greatly and $L_2$ and $L_3$ are of the same order, we can simplify our model by assuming $n$ to be constant. As the inversion charge in strong inversion mostly remains constant with the applied $V_{GS}$, we can assume $n$ to be fixed at $10^{19}$/cm³ as $n_{ch2}$ is typically $10^{19}$/cm³ in strong inversion. Now, we only need to solve (31) and (32) and find $L_2$ and $L_3$.

As shown in Fig. 2, in region R1, the slope of the surface potential decreases along the channel length and hence the shortest tunneling length $L_T$ must lie between the source and the point where the potential falls by $E_G/q$ and can be written as [28]:

$$L_T = z(\psi_{Source}) - z(\psi_{Source} - E_G/q) \qquad (34)$$

$$z(\psi_S) = L_d \cosh^{-1}((\psi_S - \psi_G)/(\psi_C - \psi_G)) \qquad (35)$$

Using $L_T$, the average electric field along the tunneling path $F_{avg}$ can be calculated as:

$$F_{avg} = E_G/qL_T \qquad (36)$$

Substituting $F_{avg}$ into the Kane's model for band to band tunneling [29], we get the expression for the drain current as:

$$I_D = A_K \times L_T \times F_{avg}^2 \times \exp(-B_K/F_{avg}) \qquad (37)$$

[23-26, 28] where $A_K$ and $B_K$ are material dependent tunnelling parameters which depend on $E_G$ and the effective mass of the carriers. The units of $A_K$ and $B_K$ are A-m/V² and V/m respectively and have to be extracted experimentally [28]. The term $A_K \times L_T$ incorporates the tunneling volume. Here, we have included the change in tunneling volume with $V_{GS}$ by taking the tunneling volume to be proportional to $L_T$. The other dimensions of the tunneling volume will be constant for a given radius of the nanowire and are part of $A_K$. The parameter $B_K$ is the exponential factor from the Kane's model for band to band tunneling and it governs the calculated values of $I_D$ [28].

In Kane's model [29], the expression for generation rate of the carriers in the tunneling region is given. This generation rate has to be integrated over the entire tunneling volume for calculating the total drain current as has been done in many of the TFET models [21, 24-26]. In our work, for simplicity, the generation rate is assumed to be constant over the entire tunneling region and hence the integration simplifies to multiplying the generation rate with tunneling volume as suggested in [28]. However, in [28] the tunneling volume is assumed to be constant for all the bias conditions and is incorporated as a part of factor $A_K$. In our model, we have modified this approach and have incorporated the change in tunneling volume with the applied gate voltage by assuming the tunneling volume to be proportional to $L_T$ and hence we get the factor $A_K \times L_T$. Therefore, the parameter $A_K$ in our work and in [28] have different mathematical units.

Most of the studies on DMG TFETs till date [15,16] suggest that $L_t$ should be much smaller than $L_a$. The study in [16] suggests that for a device of length 50 nm the optimal $L_t$ = 20 nm. Therefore, let us extend our model for a more general case where $L_t$ is much smaller than $L_a$. In such a case, the entire length of the channel under the tunneling gate may be depleted i.e. regions R1 and R2 merge into each other. This will happen more so at low values of gate voltages where $L_1$ and $L_2$ are larger. In a structure where $L_t$ is small typically below 20 nm we will first solve (22) to (32) and check if $L_1+L_2$ is larger than $L_t$. If it is true then we modify our surface potential models. Now, we have only two depletion regions: R1 which is the entire region under the tunneling gate and R3 which is the same as earlier. We will now have two equations like (17) and six unknowns ($C_1$, $D_1$, $L_1$, $C_3$, $D_3$ and $L_3$). We will again use six boundary conditions as earlier defining z = 0 as the junction of the two gates.

The tunneling gate length $L_t$ is the length of region R1 now and not $L_1$. Since regions R1 and R2 have merged, we will get a point of minimum in the surface potential at $z = -L_1$. As a result, the condition given by (23) will be different now and the value of $\psi_{s1}$ at $z = -L_t$ will be $V_{bi}$

$$\psi_{s1}(-L_t) = V_{bi} \qquad (38)$$

The other five boundary conditions given by (24)-(28) remain the same but the variables and constants of region R2 are replaced by those of region R1 (i.e. $\psi_{s2}$ will become $\psi_{s1}$ and so on). Solving as done earlier, we get the following:

$$C_3 = D_3 = \frac{\psi_{Ca} - \psi_{Ga} + \frac{q(N_S + n)L_d^2}{\varepsilon_{Si}}}{2} \qquad (39)$$

$$C_1 = D_1 = (V_{bi} - \psi_{Gt} + \frac{qN_S L_d^2}{\varepsilon_{Si}})/(2\cosh(\frac{-L_t + L_1}{L_d})) \qquad (40)$$

$$\frac{V_{bi} - \psi_{Gt} + \frac{qN_S L_d^2}{\varepsilon_{Si}}}{2\cosh(\frac{-L_t + L_1}{L_d})} \times (e^{\frac{L_1}{L_d}} + e^{\frac{-L_1}{L_d}}) + \psi_{Gt} - \frac{qN_S L_d^2}{\varepsilon_{Si}} = \qquad (41)$$

$$\frac{\psi_{Ca} - \psi_{Ga} + \frac{q(N_S + n)L_d^2}{\varepsilon_{Si}}}{2} \times (e^{\frac{L_1}{L_d}} + e^{\frac{-L_1}{L_d}}) + \psi_{Ga} - \frac{q(N_S + n)L_d^2}{\varepsilon_{Si}}$$

$$\frac{V_{bi} - \psi_{Gt} + \frac{qN_S L_d^2}{\varepsilon_{Si}}}{2\cosh(\frac{-L_t + L_1}{L_d})} \times (e^{\frac{L_1}{L_d}} - e^{\frac{-L_1}{L_d}}) = \frac{\psi_{Ca} - \psi_{Ga} + \frac{q(N_S + n)L_d^2}{\varepsilon_{Si}}}{2} \times (e^{\frac{-L_1}{L_d}} - e^{\frac{L_1}{L_d}})$$

$$(42)$$

Simultaneously solving (40)-(42) gives us $L_1$, $L_3$ and $C_1$. The drain current $I_D$ is then calculated using (34)-(37).

## III. MODEL RESULTS

The proposed models of DMG GAA nanowire TFET are verified against three dimensional numerical simulations [30]. In our simulations, we have used the models for concentration dependent mobility, electric field dependent mobility, SRH recombination, auger recombination, band gap narrowing and Kane's band to band tunneling. The parameters for Kane's band-to-band tunneling in Silvaco Atlas, $A_{Kane}$ (= 4x10$^{19}$ eV$^{1/2}$/cm-s-V$^2$) and $B_{Kane}$ (= 41 MV/cm-eV$^{3/2}$) [30] are extracted by reproducing the experimental results given in Fig. 6(a) of [28] as shown in Fig. 5. With these tunneling parameters, we simulate the device structure as shown in Fig. 1 and compare the simulation results with those predicted by our models. Fig. 6 shows the surface potential along the channel given by the model equations (22)-(32) for different values of $V_{GS}$ and $V_{DS}$ and compares them with simulation results. Fig. 7 shows the log($I_{DS}$) -$V_{GS}$ characteristic given by the model equations (34)-(37) for different values of $V_{DS}$ and compares them with simulation results. The model results are in good agreement with the simulations over a large range of $V_{GS}$ above the threshold (which is - 1.3 V). From the model results in Fig.7, it may appear unreasonable to have a lower drain current at a higher absolute drain voltage. However, this is due to the fact that our model is accurate only for gate voltages above threshold. This is because our model takes the shortest tunneling length and uses it over the entire tunneling volume. For gate voltages above the threshold, the shortest tunneling length being small dominates the tunneling current. But for gate voltages in the subthreshold region, the shortest tunneling length is large and is not able dominate the tunneling current. Hence, in the subthreshold region, using a constant tunneling length over the entire tunneling volume gives us inaccurate results. To model the drain current more accurately in the subthreshold region, a numerical approach has to be taken which integrates the tunneling current over all the 1-D tunneling lengths. Our approach to find the tunneling current is analytical and hence inaccurate in the subthreshold region. Fig. 8 shows the $I_{DS}$ -$V_{DS}$ characteristic given by the models equations (34)-(37) for different values of $V_{GS}$ and compares them with simulation results. The model results match with the simulation results with reasonable accuracy. In Fig. 9, we have plotted the surface potential along the channel for a DMG GAA nanowire TFET with $L_t$ = 20 nm and $L_a$ = 180 nm for different values of $V_{GS}$ and compared them with simulation results.

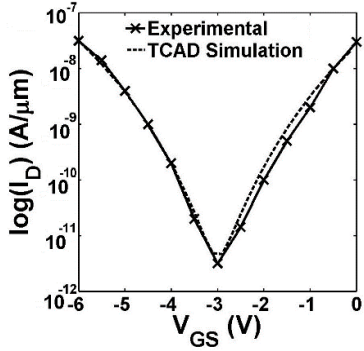

Fig. 5. Reproduction of experimental results published in Fig. 6(a) of [28] using TCAD simulations for extracting the tunneling parameters.

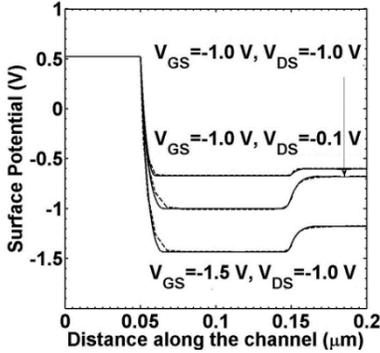

Fig. 6. Surface potential in the channel given by TCAD simulations (dashed lines) and our model (solid lines) for three biasing cases.

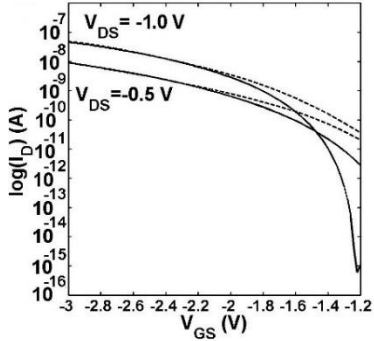

Fig. 7. log($I_D$)-$V_{GS}$ curves for the DMG GAA nanowire TFET obtained by TCAD simulations (dashed lines) and our model (solid lines).

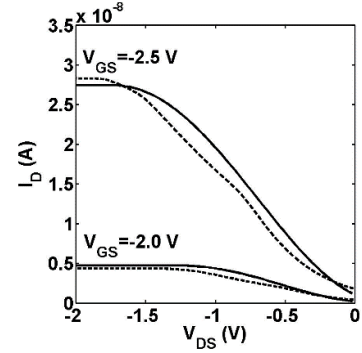

Fig. 8. $I_D$-$V_{DS}$ curves for the DMG GAA nanowire TFET obtained by TCAD simulations (dashed lines) and our model (solid lines).

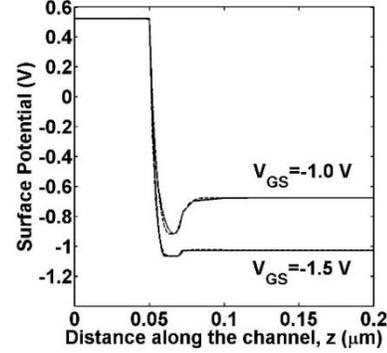

Fig. 9. Surface potential profile in the channel obtained from simulations (dashed lines) with $L_t$ = 20 nm and $L_a$ = 180 nm and model (solid lines) for $V_{DS}$ = -0.5 V and two low values of $V_{GS}$.

## IV. CONCLUSION

In this work, we have developed a pseudo 2D-analytical model for the drain current and surface potential of a DMG GAA nanowire TFET. The accuracy of our model has been tested against results obtained from three dimensional numerical simulations calibrated with experimental results. Our model can be used for a better understanding of the DMG GAA nanowire TFET and also for circuit design. It may be pointed out that our model results in the off-state and the sub-threshold region do not match with the simulations. Therefore, further work needs to be done by considering non-local tunneling to refine the models in the low current regime. Also, the models developed in this work are for a long channel device and can be used as long as all the depletion regions (i.e. regions R1, R2, R3 and R4) do not merge into each other. For a highly scaled down TFET, if these depletion regions merge with each other, a fully depleted channel approach [31] needs to be used to develop a suitable model.

## REFERENCES

[1] A. C. Seabaugh and Q. Zhang, "Low-voltage tunnel transistors for beyond CMOS logic," *Proc. IEEE*, vol. 98, no. 12, pp. 2095–2110, Dec. 2010.

[2] S. O. Koswatta, M. S. Lundstrom, and D. E. Nikonov, "Performance comparison between p-i-n tunneling transistors and conventional MOSFETs," *IEEE Trans. Electron Devices*, Vol. 56, No. 3, pp. 456–465, Mar. 2009.

[3] E.-H. Toh, G. H. Wang, G. Samudra and Y. C. Yeo, "Device physics and design of germanium tunneling field-effect transistor with source and drain engineering for low power and high performance applications", *J. Appl. Phys.*, Vol.103, No.10, pp. 104504 - 104504-5, 2008.

[4] M. J. Kumar and S. Janardhanan, "Doping-less Tunnel Field Effect Transistor: Design and Investigation", *IEEE Trans. on Electron Devices*, Vol.60, pp.3285 - 3290, October 2013.

[5] S. Saurabh and M. J. Kumar, "Estimation and Compensation of Process Induced Variations in Nanoscale Tunnel Field Effect Transistors (TFETs) for Improved Reliability," *IEEE Trans. on Device and Materials Reliability*, Vol.10, pp.390 - 395, September 2010.

[6] S. Saurabh and M. J. Kumar, "Impact of Strain on Drain Current and Threshold Voltage of Nanoscale Double Gate Tunnel Field Effect Transistor (TFET): Theoretical Investigation and Analysis," *Japanese Journal of Applied Physics*, Vol.48, paper no. 064503, June 2009.

[7] International Technology Roadmap for Semiconductor, http://www.itrs.net/.

[8] C. Shen, S.-L. Ong, C.-H. Heng, G. Samudra, and Y.-C. Yeo, "A Variational Approach to the Two-Dimensional Nonlinear Poisson's Equation for the Modeling of Tunneling Transistors", *IEEE Electron Device Lett.*, Vol.29, No.11, pp.1252-1255, November 2008.

[9] K. Boucart and A. M. Ionescu, "A new definition of threshold voltage in Tunnel FETs," *Solid State Electron.*, Vol. 52, No. 9, pp. 1318–1323, Sep. 2008.

[10] Q. Shao, C. Zhao, C. Wu, J. Zhang, L. Zhang and Z. Yu, "Compact model and Projection of Silicon Nanowire Tunneling Transistors (NWTFETs)", in *Int. Conf. of Elec. Dev. and Solid-State Circuits (EDSSC)*, June 2013, pp.1-2.

[11] K. H. Yeo, Y. S. D. Suk, M. Li, Y.-Y. Yeoh, K. H. Cho, K.-H. Hong, S. Yun, M. S. Lee, N. Cho, K. Lee, D. Hwang, B. Park, D.-W. Kim, D. Park and B.-I. Ryu, "Gate-all-around (GAA) twin silicon nanowire MOSFET (TSNWFET) with 15 nm length gate and 4 nm radius nanowire," *Proc. IEDM*, Dec. 2006, pp. 539–542.

[12] H.S.P Wong, "Beyond the conventional transistor," *IBM Jl. of Research & Development*, vol. 46, no. 2/3, pp.133-168, March/May 2002.

[13] N. Jain, E. Tutuc, S.K. Banerjee and L.F. Register, "Performance Analysis of Germanium Nanowire Tunneling Field Effect Transistors", in *Device Research Conference*, June 2008, pp.99-100.

[14] K. Jeon, "Si Tunnel Transistors with a Novel Silicided Source and 46mV/dec Swing", in *Symp. on VLSI Tech. Digest*, 2010, pp.121-122.

[15] A. Zhan, J. Mei, L. Zhang, H. He, J. He and M. Chan, "Numerical Study on Dual Material Gate Nanowire Tunnel Field-Effect Transistor" in *Int. Conf. on Elec. Devices and Solid State Circuit (EDSSC)*, 2012, pp.1-5.

[16] S. Saurabh and M. J. Kumar, "Novel Attributes of a Dual Material Gate Nanoscale Tunnel Field-Effect Transistor", *IEEE Trans. Electron Devices*, Vol.58, No.2, pp.404-410, February 2011.

[17] N. Cui, R. Liang, and J. Xu, "Heteromaterial gate tunnel field effect transistor with lateral energy band profile modulation", *Appl. Phys. Letts.*, Vol. 98, pp. 142105-142105-3, 2011.

[18] H. Lou, L. Zhang, Y. Zhu, X. Lin, S. Yang, J. He, and M. Chan, "A Junctionless Nanowire Transistor With a Dual-Material Gate", *IEEE Trans. Electron Devices*, Vol.59, No.7, pp.1829-1836, July 2012.

[19] M. J. Lee and W. Y. Choi, "Effects of Device Geometry on Hetero-Gate-Dielectric Tunneling Field-Effect Transistors", *IEEE Electron Device Lett.*, Vol.33, No.10 pp.1459-1461, October 2012.

[20] M.-L. Fan, V. P. Hu, Y.-N. Chen, P. Su and C.-T. Chuang, "Analysis of Single-Trap-Induced Random Telegraph Noise and its Interaction With Work Function Variation for Tunnel FET", *IEEE Trans. Electron Devices*, Vol.60, No.6, pp.2038-2044, June 2013.

[21] A. S. Verhulst, B. Sorée, D. Leonelli, W. G. Vandenberghe and G. Groeseneken, "Modeling the single-gate, double-gate, and gate-all-around tunnel field-effect transistor", *J. App. Phys.*, Vol.107, paper no. 024518, June 2010.

[22] B. Bhushan, K. Nayak, and V. R. Rao, "DC Compact Model for SOI Tunnel Field-Effect Transistors", *IEEE Trans. Electron Devices*, Vol.59, No.10, pp.2635-2642, October 2012.

[23] W. G. Vandenberghe, A. S. Verhulst, G. Groeseneken, B. Soree, and W. Magnus, "Analytical Model for a Tunnel Field-Effect Transistor" *IEEE Mediterranean Electrotechnical Conf.*, pp. 923-928, 5-7 May 2008.

[24] N. Cui, R. Liang, J. Wang and J. Xu, "Si-based Hetero-Material-Gate Tunnel Field Effect Transistor: Analytical Model and Simulation", *IEEE Conf. on Nanotechnology (IEEE-NANO)*, pp. 1-5, Aug 2012.

[25] W. G. Vandenberghe, A. S. Verhulst, G. Groeseneken, B. Soree and W. Magnus, "Analytical Model for Point and Line Tunneling in a Tunnel Field-Effect Transistor", *Int. Conf. on Simulation of Semiconductor Processes and Devices (SISPAD)*, pp. 137-140, Sep 2008.

[26] E-H Toh, G. H. Wang, G. Samudra, and Y-C. Yeo, "Device physics and design of double-gate tunneling field-effect transistor by silicon film thickness optimization", *Appl. Phys. Lett.*, Vol.90, pp. 263507-263507-3, 2007.

[27] M. G. Bardon, H. P. Neves, R. Puers and C. V. Hoof, "Pseudo-Two-Dimensional Model for Double-Gate Tunnel FETs Considering the Junctions Depletion Regions", *IEEE Trans. Electron Devices*, Vol.57, No.4, pp.827-834, April 2010.

[28] J. Wan, C. L. Royer, A. Zaslavsky and S. Cristoloveanu, "A tunnelling field effect transistor model combining interband tunneling with channel transport", *J. Appl. Phys.*, Vol.110, No.10, pp.104503 - 104503-7, 2011.

[29] E. O. Kane, "Zener tunneling in semiconductors," *J. Phys. Chem. Solids*, Vol. 12, pp. 181-188, 1959.

[30] ATLAS Device Simulation Software, *Silvaco Int.*, Santa Clara, CA, 2012.

[31] M. J. Kumar and A. Chaudhry, "Two-Dimensional Analytical Modeling of Fully Depleted DMG SOI MOSFET and Evidence for Diminished SCEs", *IEEE Trans. on Electron Devices*, Vol. 51, No. 4, pp. 569-574, Apr 2004.

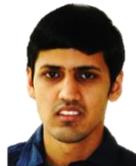

**Rajat Vishnoi** received the B.Tech. degree in Electrical engineering from Indian Institute of Technology, Kanpur, India, in 2012. He is currently pursuing the PhD degree at the Indian Institute of Technology, Delhi, India. His research interests include semiconductor device modelling, design and fabrication.

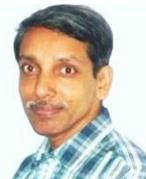

**M. Jagadesh Kumar** is the NXP (Philips) Chair Professor established at IIT Delhi by Philips Semiconductors, The Netherlands. He is also a Principal Investigator of the Nano-scale Research Facility at IIT Delhi. He is an Editor of the IEEE TRANSACTIONS ON ELECTRON DEVICES. For more details on Dr. Jagadesh Kumar, please visit http://web.iitd.ac.in/~mamidala